\documentclass[12pt]{article}
\usepackage[bookmarks,bookmarksnumbered]{hyperref}
\usepackage{amsmath,XE}
\usepackage{graphicx,color,epstopdf}
\usepackage{booktabs,multirow}

\def\sC{\mathscr{C}}
\def\ddt{\partial_\tau}
\def\wt{\mathop\text{wt}\nolimits}

\def\hi{{\hat\imath}}
\def\hj{{\hat\jmath}}
\def\vC#1{\vcenter{\hbox{\hss#1\hss}}}
\def\bs#1{\boldsymbol{#1}}
\definecolor{WIP}{rgb}{.8,0,.2}

\definecolor{Green}  {rgb}{0.10,0.61,0.22}
\definecolor{Orange} {rgb}{0.83,0.51,0.23}
\definecolor{Red}    {rgb}{0.78,0.00,0.12}
\definecolor{Purple} {rgb}{0.42,0.15,0.45}
\definecolor{Turque} {rgb}{0.00,0.63,0.85}
\definecolor{Blue}   {rgb}{0.00,0.00,1.00}
\definecolor{Magenta}{rgb}{1.00,0.00,1.00}
\definecolor{Gold}   {rgb}{.90,0.79,0.02} 
\definecolor{Seaweed}{rgb}{0.01,0.24,0.09}
\definecolor{Brown}  {rgb}{0.43,0.26,0.32}
\definecolor{Grey}  {rgb}{0.40,0.40,0.40}
\def\C#1#2{{\ifcase#1\or
             \color{Green}\or \color{Orange}\or \color{Red}\or
              \color{Purple}\or \color{Turque}\or \color{Blue}\or
               \color{Magenta}\or \color{Gold}\or \color{Seaweed}\or
                \color{Brown}\else\color{Grey}\fi#2}}
\def\Ft#1{\,\footnote{#1}}
\newdimen\parshift\parshift=\parindent
\catcode`@=11
 \long\def\@footnotetext#1{\insert\footins{\reset@font\footnotesize\interlinepenalty%
  \interfootnotelinepenalty\splittopskip\footnotesep\splitmaxdepth\dp\strutbox%
   \floatingpenalty\@MM\hsize\columnwidth\addtolength{\hsize}{-2\parindent}
    \@parboxrestore\protected@edef\@currentlabel{\csname p@footnote\endcsname\@thefnmark}
      \color@begingroup
       \@makefntext{\rule\z@\footnotesep\ignorespaces#1\@finalstrut\strutbox}
        \color@endgroup}}
 \long\def\@makefntext#1{\hglue\parshift
                         \vbox{\noindent\hb@xt@0em{\hss\@makefnmark}#1}}
\catcode`@=12
 \font\rOpe=cmsy10                        
 \def\ktl{{\hbox{\rOpe\char'170}}}        
 \def\kbl{{\hbox{\rOpe\char'170}}}        
 \def\kcr{{\reflectbox{\rOpe\char'170}}}        
 \def\ktr{{\reflectbox{\rOpe\char'170}}}        
 \def\kbr{{\reflectbox{\rOpe\char'170}}}        
 \def\Border{\vbox{\hsize0pt
        \setlength{\unitlength}{1mm}
        \newcount\xco
        \newcount\yco
        \xco=-21
        \yco=12
        \begin{picture}(0,0)(-7.5,0)
        \put(\xco,\yco){$\ktl$}
        \advance\yco by-1
        {\loop
        \put(\xco,\yco){$\kcr$}
        \advance\yco by-2
        \ifnum\yco>-240
        \repeat
        \put(\xco,\yco){$\kbl$}}
        \xco=170
        \yco=12
        \put(\xco,\yco){$\ktr$}
        \advance\yco by-1
        {\loop
        \put(\xco,\yco){$\kcr$}
        \advance\yco by-2
        \ifnum\yco>-240
        \repeat
        \put(\xco,\yco){$\kbr$}}
        \put(-19.5,13){\scalebox{.54}{State University of New York
            Physics Department|University of Maryland Center for
            String and Particle  Theory \&\ Physics Department|%
            Howard University Physics \&\ Astronomy Department}}
        \put(-19.5,-241.5){\scalebox{.649}{University of Alberta Mathematical
            and Statistical Sciences Department|Pepperdine University Natural
            Sciences Division|Bard College Mathematics Program}}
        \end{picture}
        \par\vskip-8mm}}
\definecolor{UMred}{rgb}{.9,.05,.2}
 \def\UMbanner{\vbox{\hsize0pt
        \setlength{\unitlength}{.4mm}
        \thicklines
        \begin{picture}(0,0)(-30,-10)
        \put(165,16){\line(1,0){4}}
        \put(170,16){\line(1,0){4}}
        \put(180,16){\line(1,0){4}}
        \put(175,0){\line(1,0){4}}
        \put(180,0){\line(1,0){4}}
        \put(185,0){\line(1,0){4}}
        \put(169,0){\line(0,1){16}}
        \put(170,0){\line(0,1){16}}
        \put(179,0){\line(0,1){16}}
        \put(180,0){\line(0,1){16}}
        \put(184,0){\line(0,1){16}}
        \put(185,0){\line(0,1){16}}
        \put(169,16){\oval(8,32)[bl]}
        \put(170,16){\oval(8,32)[br]}
        \put(179,0){\oval(8,32)[tl]}
        \put(185,0){\oval(8,32)[tr]}
        \end{picture}
        \par\vskip-6.5mm
        \thicklines}}

 \SfTitles 
 \allowdisplaybreaks
 \seceq
\begin{document}
\thispagestyle{empty}
\vbox{\Border\UMbanner}
 \noindent
 \today\hfill UMDEPP~09-029\\[-1mm]\hglue0mm\hfill SUNY-O/672
  \vfill
 \begin{center}
{\LARGE\sf\bfseries A Superfield for Every Dash-Chromotopology
 }\\*[5mm]
{\sf\bfseries C.F.\,Doran$^a$,
              M.G.\,Faux$^b$,
              S.J.\,Gates, Jr.$^c$,
              T.\,H\"{u}bsch$^d$,
              K.M.\,Iga$^e$
              and
              G.D.\,Landweber$^f$}\\*[2mm]
{\small\it
  $^a$Department of Mathematical and Statistical Sciences,\\[-1mm]
      University of Alberta, Edmonton, Alberta, T6G 2G1 Canada%
  \\[-4pt] {\tt  doran@math.ualberta.ca}
  \\
  $^b$Department of Physics,\\[-1mm]
      State University of New York, Oneonta, NY 13825%
  \\[-4pt] {\tt  fauxmg@oneonta.edu}
  \\
  $^c$Center for String and Particle Theory,\\[-1mm]
      Department of Physics, University of Maryland, College Park, MD 20472%
  \\[-4pt] {\tt  gatess@wam.umd.edu}
  \\
  $^d$Department of Physics \&\ Astronomy,\\[-1mm]
      Howard University, Washington, DC 20059
  \\[-4pt] {\tt  thubsch@howard.edu}
  \\
  $^e$Natural Science Division,\\[-1mm]
      Pepperdine University, Malibu, CA 90263%
  \\[-4pt] {\tt  Kevin.Iga@pepperdine.edu}
  \\
 $^f$Department of Mathematics, Bard College,\\[-1mm]
     Annandale-on-Hudson, NY 12504-5000%
  \\[-4pt] {\tt  gregland@bard.edu}
 }\\[5mm]
  \vfill
{\sf\bfseries ABSTRACT}\\[2mm]
\parbox{117mm}{
The recent classification scheme of so-called adinkraic off-shell supermultiplets of $N$-extended worldline supersymmetry without central charges finds a combinatorial explosion. Completing our earlier efforts, we now complete the constructive proof that all of these trillions or more of supermultiplets have a superfield representation. While different as superfields and supermultiplets, these are still super-differentially related to a much more modest number of minimal supermultiplets, which we construct herein.}
 \vfill
\end{center}
 \vfill
\noindent PACS: 11.30.Pb, 12.60.Jv

\clearpage\setcounter{page}{1}
\section{Introduction}
 \label{int}
The $N$-extended supersymmetry on the worldline and without central charges is defined  by:
\begin{equation}
 \begin{gathered}
  \{Q_I,Q_J\} = 2\,\delta_{IJ}\,H,\qquad
  [H,Q_I]=0,\qquad I,J=1,\cdots,N,\\
  (Q_I)^\dagger=Q_I,\qquad (H)^\dagger = H,
 \end{gathered}
 \label{eSuSy}
\end{equation}
where $H$ is the worldline Hamiltonian, identifiable with $i\hbar\,\ddt$, and $Q_I$ is the $I^\text{th}$ supercharge. Physical interest in this algebra stems from three separate and logically independent applications:
\begin{enumerate}\itemsep=-3pt\vspace{-3mm}
 \item Dimensional reduction of any supersymmetric theory in ``actual'' spacetime: supersymmetric Yang-Mills gauge theories, the supersymmetric Standard Model of particle physics, {\em\/etc.\/};
 \item The {\em\/underlying\/} description or dimensional reduction thereof, in theories of extended objects, such as the worldsheet description of superstring theory, or the matrix version of $M$-theory;
 \item Induced supersymmetry in the Hilbert space of a supersymmetric theory, in the  Schr\"odinger picture; $H,Q_I$ are expressed in terms of particle state creation and annihilation operators.
\end{enumerate}\vspace{-3mm}
While not limited in principle, $N\leq32$ seems to suffice in all known fundamental physics.

Although Eqs.\eq{eSuSy} are covariant with respect to an $\textsl{O}(N)$ symmetry, under which the $Q_I$ span the vector representation, we assume no part of any symmetry, other than $N$-extended supersymmetry itself. On occasion, such as in\eq{eSysD4} or\eq{eH8}, the full $\textsl{O}(N)$ will indeed turn out to be a symmetry; in other cases, such as in\eq{eD6}, this symmetry will be explicitly broken to a subgroup: in\eq{eD6}, $\textsl{O}(6)\to\textsl{O}(2)^{\otimes3}$. As usual, insisting on the {\em\/least\/} amount of symmetry provides for the {\em\/most\/} generality; imposing symmetries will narrow down our results.

The classification of off-shell supermultiplets of the algebra\eq{eSuSy} has remained an open problem for over three decades.
 Focusing on the worldline ``shadow'' of supersymmetric theories in higher dimensional spacetime avoids all technical and notational difficulties related to the Lorentz symmetry in actual, higher-dimensional spacetimes. Lorentz and other symmetry considerations can be treated as ``internal'', unrelated to spacetime, and can be included subsequently in the reverse of the dimensional reduction, the oxidization of Ref.\cite{rGR0}.
 In this vein, Refs.~\cite{rGR0,rGR-1,rGR1,rGR2,rGR3,rGLPR,rGLP} and then~\cite{rA,r6-1,r6--1,r6-3c,r6-3,r6-3.2,r6-3.4} forged a novel approach, employing graph theory and error-correcting codes, which resulted in a combinatorially growing number of {\em\/Adinkras\/}|graphs that represent each supermultiplet.
 Application of these techniques to concrete and previously unsolved problems in supersymmetric physics was demonstrated in Ref.~\cite{r6-2,r6-3a,r6-4,r6-7a,r6-4.2}. Ref.\cite{r6-1} also begun a rigorous translation between these novel, {\em\/adinkraic\/} results into the much more standard methods of superspace\cite{r1001,rYM84New,rPW,rWB,rYM97Gau,rBK}.

The purpose of this note is to complete the translation of the results of this adinkraic classification scheme\cite{r6-1,r6-3,r6-3.2,r6-3.4} into superspace, begun in Ref.\cite{r6-1}. To that end,  Section~\ref{s:R} briefly reviews these results, the so-obtained classification scheme, and the part of the translation known this far. In particular, Ref.\cite{r6-1} ends with a conjecture that we are now able to prove, in Section~\ref{10*12}, owing in part to the subsequent developments\cite{r6-3,r6-3.2}.
Section~\ref{s:C} collects a couple of clarifying examples and a few concluding comments.

\section{Adinkraic Results and Translation into Superspace}
 \label{s:R}
The adinkraic classification scheme of Refs.\cite{r6-1,r6-3,r6-3.2} focuses on {\em\/adinkraic supermultiplets\/}. These consist of bosons $\f_i(\t)$ and fermions $\j_\hi(\t)$, and supersymmetry acts amongst these so that for any fixed $Q_I$ and $\f_i(\t)$,
\begin{equation}
 Q_I\, \f_i(\t) = \pm\ddt^\l\, \j_\hi(\t),
 \qquad \l=0,1,
   \label{eQb}
\end{equation}
for some definite fermionic component field, and conversely
\begin{equation}
 Q_I\,  \j_\hi(\t) = \pm i\,\ddt^{1-\l}\, \f_i(\t).
 \label{eQf}
\end{equation}
The structure of an adinkraic supermultiplet may be faithfully depicted by an {\em\/Adinkra\/}:
 ({\small\bf1})~Assign a node to every component field: white for bosons and black for fermions.
 ({\small\bf2})~Draw an edge in the $I^\text{th}$ color from node $v_1$ to node $v_2$ precisely if the component field $F_2$ of $v_2$ is the $Q_I$-image of the component field $F_1$ of $v_1$ and $[F_2]=[F_1]+\frac12$, where $[F]$ is the engineering unit of $F$.
 ({\small\bf3})~An edge is drawn solid for the choice of ``$+$'' in Eqs.~(\ref{eQb})--(\ref{eQf}), and dashed for the ``$-$'' choice.
 See Table~\ref{t:A} for a dictionary.
\begin{table}[htbp]
  \centering\setlength{\unitlength}{1mm}
  \begin{tabular}{@{} cc|cc @{}}
    \makebox[15mm]{\sf\bfseries Adinkra}
  & \makebox[40mm]{\sf\bfseries\boldmath$Q$-action} 
  & \makebox[15mm]{\sf\bfseries Adinkra}
  & \makebox[40mm]{\sf\bfseries\boldmath$Q$-action} \\ 
    \hline
    \begin{picture}(5,9)(0,5)
     \put(0,0){\includegraphics[height=11mm]{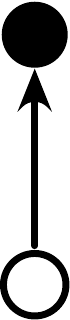}}
     \put(3,0){\scriptsize$i$}
     \put(3,9){\scriptsize$\hi$}
     \put(-1,4){\scriptsize$I$}
    \end{picture}\vrule depth4mm width0mm
     & $Q_I\begin{bmatrix}\psi_\hi\\\phi_i\end{bmatrix}
           =\begin{bmatrix}i\dot\phi_i\\\psi_\hi\end{bmatrix}$
  & \begin{picture}(5,9)(0,5)
     \put(0,0){\includegraphics[height=11mm]{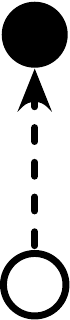}}
     \put(3,0){\scriptsize$i$}
     \put(3,9){\scriptsize$\hi$}
     \put(-1,4){\scriptsize$I$}
    \end{picture}\vrule depth4mm width0mm
     & $Q_I\begin{bmatrix}\psi_\hi\\\phi_i\end{bmatrix}
           =\begin{bmatrix}-i\dot\phi_i\\-\psi_\hi\end{bmatrix}$ \\[5mm]
    \hline
    \begin{picture}(5,9)(0,5)
     \put(0,0){\includegraphics[height=11mm]{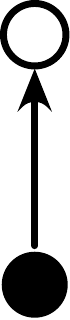}}
     \put(3,0){\scriptsize$\hi$}
     \put(3,9){\scriptsize$i$}
     \put(-1,4){\scriptsize$I$}
    \end{picture}\vrule depth4mm width0mm
     &  $Q_I\begin{bmatrix}\phi_i\\\psi_\hi\end{bmatrix}
           =\begin{bmatrix}\dot\psi_\hi\\i\phi_i\end{bmatrix}$
  & \begin{picture}(5,9)(0,5)
     \put(0,0){\includegraphics[height=11mm]{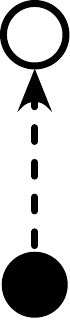}}
     \put(3,0){\scriptsize$\hi$}
     \put(3,9){\scriptsize$i$}
     \put(-1,4){\scriptsize$I$}
    \end{picture}\vrule depth4mm width0mm
     &  $Q_I\begin{bmatrix}\phi_i\\\psi_\hi\end{bmatrix}
           =\begin{bmatrix}-\dot\psi_\hi\\-i\phi_i\end{bmatrix}$ \\[5mm]
    \hline
  \multicolumn{4}{l}{\vrule height3.5ex width0pt\parbox{120mm}{\small\baselineskip=12pt The edges are here labeled by the variable index $I$; for any fixed $I$, each corresponding edge is drawn in the $I^{\text{th}}$ color instead.}}
  \end{tabular}
  \caption{\baselineskip=13pt
  The correspondences between the Adinkra components and supersymmetry transformation formulae~(\ref{eQb})--(\ref{eQf}):
    vertices\,$\leftrightarrow$\,component fields;
    vertex color\,$\leftrightarrow$\,fermion/boson;
    edge color/index\,$\leftrightarrow$\,$Q_I$;
    edge dashed\,$\leftrightarrow$\,``$-$'' in\eq{eQb}; and
    orientation\,$\leftrightarrow$\,placement of $\partial_\tau$.
    They apply to all $\phi_A,\psi_B$ within a supermultiplet and all $Q_I$-transformations amongst them.}
  \label{t:A}
\end{table}
For clarity, we dispense with the arrows on the edges, but position the nodes so that all edges are oriented upward, and each node is placed at a height that is proportional to the engineering unit of the corresponding component field~\cite{r6-1}.

The connectivity between component fields provides a notion of topology to every supermultiplet; since edges corresponding to distinct $Q_I$'s are drawn in distinct colors and dashed for ``$-$'' in\eq{eQb}, the topology including this information is called the {\em\/dash-chromotopology\/} of the Adinkra and of the corresponding supermultiplet. 

Ref.\cite{r6-1} then partitions the representations of $N$-extended worldline supersymmetry without central charges into ``families'' of Adinkras, wherein all members have the same dash-chromotopology, but differ in ``hanging''. For example,
\begin{equation}
 \vC{\includegraphics[width=30mm]{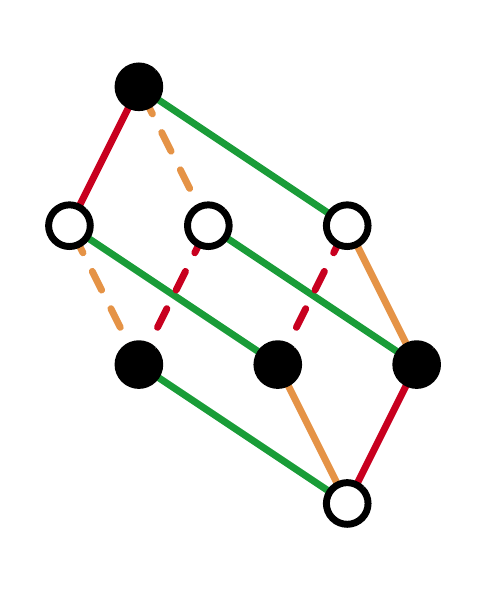}}
 \vC{\includegraphics[width=30mm]{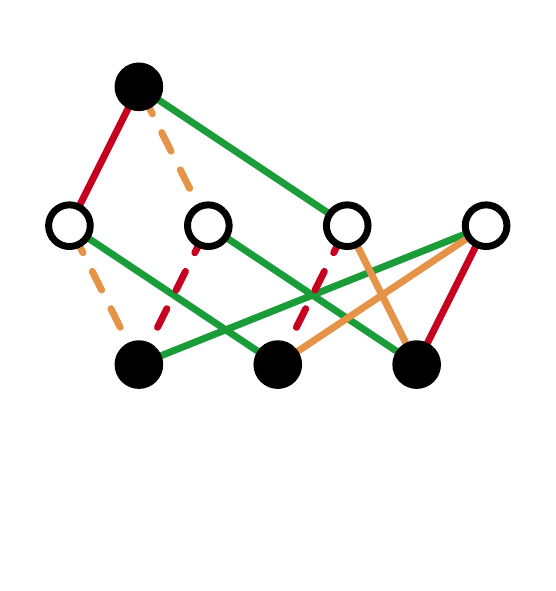}}
 \vC{\includegraphics[width=30mm]{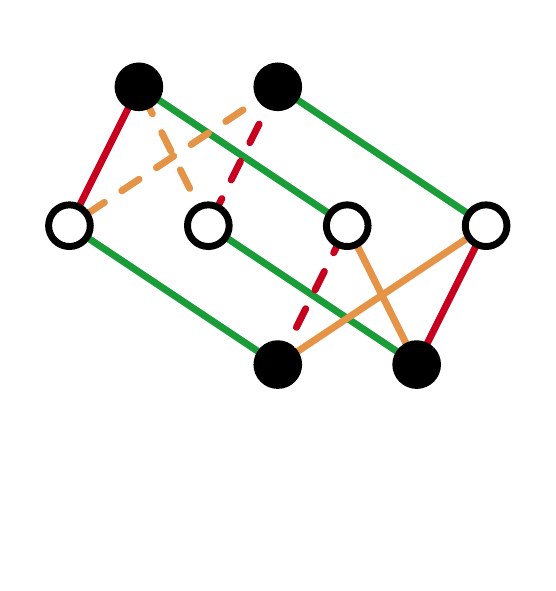}}
 \vC{\includegraphics[width=30mm]{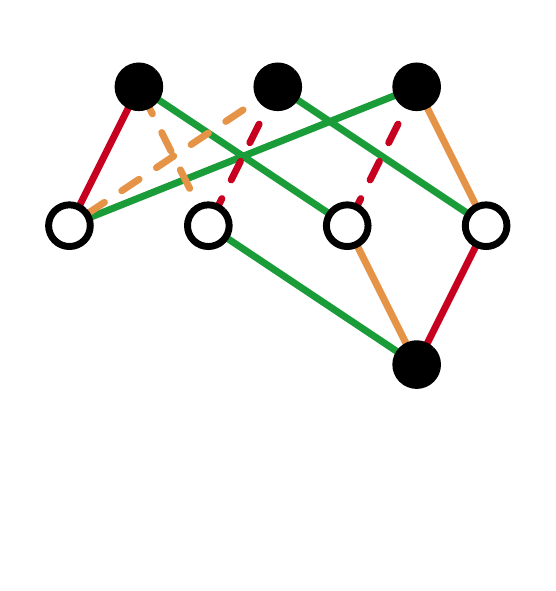}}
 \vC{\includegraphics[width=30mm]{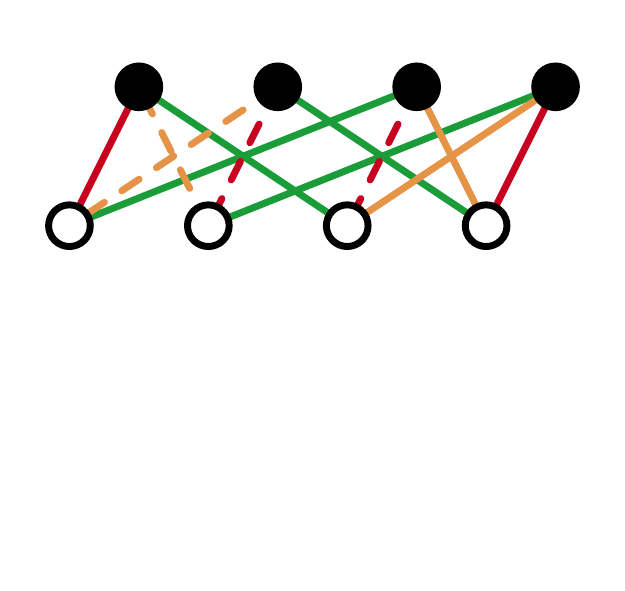}}
 \label{eN3Seq}
\end{equation}
are some of the $N=3$ Adinkras; they all have the same dash-chromotopology, equal to the 3-cube with the indicated edges dashed\Ft{Distinct choices of edge-dashing may well be equivalent by a sign-redefinition on some of the component fields, and so form equivalence classes. The classification of these equivalence classes and a homology computation that identifies to which particular equivalence class does a given Adinkra belong is specified in Ref.\cite{r6-3.4}.}.
 Each Adinkra in the sequence\eq{eN3Seq} is obtained from the one on the left by raising one of the nodes. Theorems~5.1 and~5.3 of Ref.\cite{r6-1} and their respective corollaries rigorously prove that all Adinkras of the same dash-chromotopology may be obtained one from another in this fashion, and that each such family contains:
 ({\small\bf1})~at least one {\em\/Valise\/}, where all bosons and all fermions are on two adjacent levels, as in the right-most Adinkra in\eq{eN3Seq},
 ({\small\bf2})~at least one maximally extended Adinkra (``top Adinkra'' in Ref.\cite{rA}) that appears to hang freely, hanged from a single highest node, such as the left-most Adinkra in\eq{eN3Seq}, and
 ({\small\bf3})~at least one maximally extended Adinkra that appears to float freely upward from a single lowest, anchoring node, such as is also the left-most Adinkra in\eq{eN3Seq}.
 Theorem~7.6 of Ref.\cite{r6-1} then proves that for every given family (dash-chromotopology) of Adinkras|if any one of its members has a superfield representation|all others can be constructed from it, following the provided algorithm. 

Refs.\cite{r6-3,r6-3.2} prove that
 ({\small\bf1})~the chromotopology of every Adinkra is $[0,1]^N/\sC$, where $\sC$ is a doubly-even linear binary block code encoding a $(\ZZ_2)^k$-action on $[0,1]^N$, and that
 ({\small\bf2})~every such quotient, $[0,1]^N/\sC$, defines an Adinkra chromotopology.
For a telegraphic review of this isomorphism, let $\sC$ be generated by the binary codewords $\bs{b}_a=(b_{a1},\cdots,b_{aN})$, each of which defines an operator:
\begin{equation}
  \bs{b}_a=(b_{a1},\cdots,b_{aN})\qquad\mapsto\qquad
  \bs{Q}^{\bs{b}_a}:=Q_1^{b_{a1}}\cdots Q_N^{b_{aN}}\qquad
  a=1,\cdots,k.
\end{equation}
$\sC$ being a doubly-even binary linear block code means that $b_{aI}\in\{0,1\}$, the number of 1's in each $\bs{b}_a$ is divisible by four, and the bitwise product of any two codewords has an even number of 1's:
\begin{equation}
 \wt(\bs{b}_a)=0\pmod4;\qquad
 \wt(\bs{b}_a):=\sum_{I=1}^Nb_{aI}\quad\text{is the {\em\/Hamming weight\/}}.
\end{equation}
These in turn imply that
 $\bs{Q}^{\bs{b}_a}$ contains every $Q_I$ at most once,
 $(\bs{Q}^{\bs{b}_a})^2=+H^{\wt(\bs{b}_a)}$ for every $a$, and
 $[\bs{Q}^{\bs{b}},\bs{Q}^{\bs{b}'}]=0$, for any two $\bs{b},\bs{b}'\in\sC$, not just the generators.

Within any adinkraic supermultiplet $\bs{M}=(\f_1,\cdots,\f_m|\j_1,\cdots,\j_m)$, such operators act:
\begin{equation}
 \bs{Q}^{\bs{b}_a}(\f_i) = c(\ddt^{\l_{aij}}\f_j),\quad\text{(no summation!)}\qquad
 \l_{aij}:= \wt(\bs{b}_a)+[\f_i]-[\f_j],
 \label{eaij}
\end{equation}
for some definite $\f_j\in\bs{M}$ on the right-hand side, some coefficient $c$, and where $[\f_i]$ denotes the engineering unit of $\f_i$. Analogous formulae for fermions define $\l_{a\hi\hj}:=\wt(\bs{b}_a)+[\j_\hi]-[\j_\hj]$.

If $\l_{aij}=\tfrac12\wt(\bs{b}_a)=\l_{a\hi\hj}$ for all $\sC$-generators $\bs{b}_a$ and all $\f_i,\j_\hi\in\bs{M}$, then $[\f_i]=[\f_j]$ for each pair of bosonic component fields associated by the relation\eq{eaij}; the analogous also holds for all fermionic pairs so connected.
In that case,
\begin{equation}
 \hat\p^\pm_a(\f_i)=\pm\,c\,\f_j
  \qquad\text{and}\qquad
 \hat\p^\pm_a(\f_j)=\pm\frac1{c}\,\f_i
 \label{e}
\end{equation}
defines for each generator, $\bs{b}_a\in\sC$ an engineering unit-preserving $\ZZ_2$-reflection symmetry within the supermultiplet.
 Corresponding to each generator $\bs{b}_a$, the projection $\f_i\mapsto(\f_i+\vs_a\,c\,\f_j)$ ``halves'' the supermultiplet; iterating this for each generator produces
\begin{equation}
  \bs{M}|_{\sC,\vec\vs},\qquad
  \vec\vs=(\vs_1,\cdots,\vs_k),\quad \vs_a=\pm1,\quad a=1,\cdots,k,
 \label{eM/C}
\end{equation}
a collection of $2^k$ quotient supermultiplets, each with $1/2^k$ component fields of the original $\bs{M}$. A simple example of this is the $\sC=d_4$ case with a single generator corresponding to $Q_1Q_2Q_3Q_4$:
\begin{equation}
 \vC{\begin{picture}(70,30)(0,2)
      \put(0,5){\includegraphics[height=28mm]{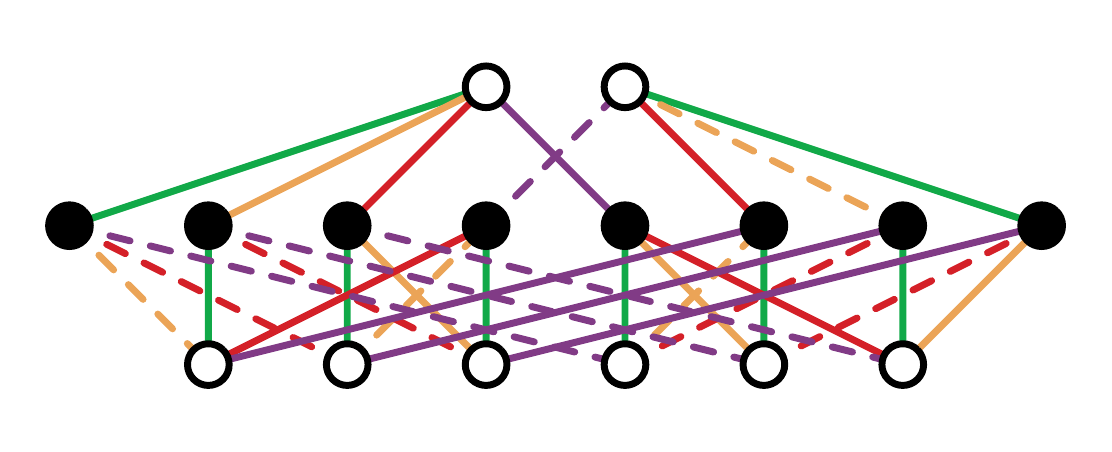}}
      \put(32,3){$\bs{M}$}
     \end{picture}}
  \buildrel{\textstyle\ZZ_2}\over\longrightarrow
 \vC{\begin{picture}(34,30)(0,2)
      \put(0,5){\includegraphics[height=28mm]{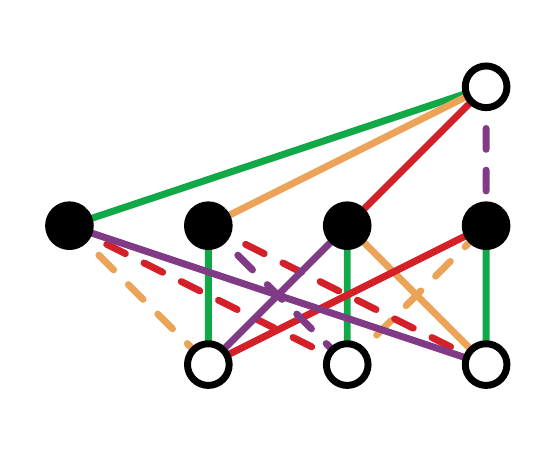}}
      \put(12,3){$\bs{M}|_{d_4,+}$}
     \end{picture}}
  ~\text{or}~
 \vC{\begin{picture}(34,30)(0,2)
      \put(0,5){\includegraphics[height=28mm]{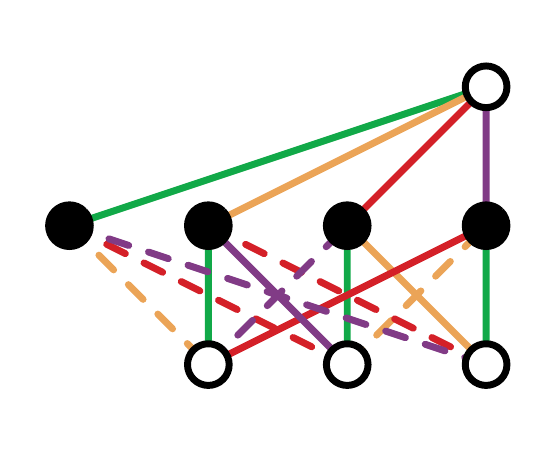}} 
      \put(12,3){$\bs{M}|_{d_4,-}$}
     \end{picture}}
 \label{eN4split}
\end{equation}
where the Adinkra on the left-hand side has the chromotopology of a 4-cube and the $\ZZ_2$ symmetry\eq{eaij} is a left-right reflection; the two Adinkras on the right-hand side are the $\vs=+1$ and $\vs=-1$ projections with respect to this symmetry. That the equality $Q_1Q_2Q_3Q_4=\pm H^2$ holds throughout the supermultiplet corresponds to the fact that in the projected Adinkras on the right-hand side of\eq{eN4split}, all 4-color quadrilaterals are closed, and moreover, the product of signs (dashedness) along each quadrilateral either equals the sign of the permutation of the colors along the quadrilateral (in $\bs{M}|_{d_4,+}$), or is opposite (in $\bs{M}|_{d_4,-})$.

Doubly-even binary linear block codes for $N\leq32$ have not all been listed so far, and Ref.\cite{r6-3} started a distributed supercomputing program, which has completed the $N\leq28$ listing and and is expected to compute some trillions of $N\leq32$ codes. Each such code corresponds to a family of Adinkras, one member of a family differing from another in how its nodes are hanged|such as those in the sequence\eq{eN3Seq}. The number of distinct hanging arrangements for the Adinkras evidently grows combinatorially with their size and so with $N$. Among these, certain pairs correspond to isomorphic supermultiplets\cite{r6-3.2}, but the total number of inequivalent supermultiplets for $N\leq32$ is still well beyond trillions: the complete list of Adinkras is beyond journal publication already for $N=5$; see Ref.\cite{r6-3.2} for more information, as well as for an algorithm for listing only the non-isomorphic supermultiplets.

By virtue of being an expansion over the exterior algebra generated by the $\q^I$, the familiar, unconstrained, real Salam-Strathdee superfield is the supermultiplet with the so-called `top' Adinkra\cite{rA}, with the chromotopology of the $N$-cube\cite{r6-1,r6-3}. These are $\big(1\big|\binom{N}1\big|\binom{N}2\big|\cdots|\binom{N}{N-1}\big|1\big)$-dimensional representations of $N$-extended supersymmetry without central charges, unique up to the choice of the spin-statistics of the lowest component:
\vspace{-5mm}
\begin{equation}
 \vC{\includegraphics[width=160mm]{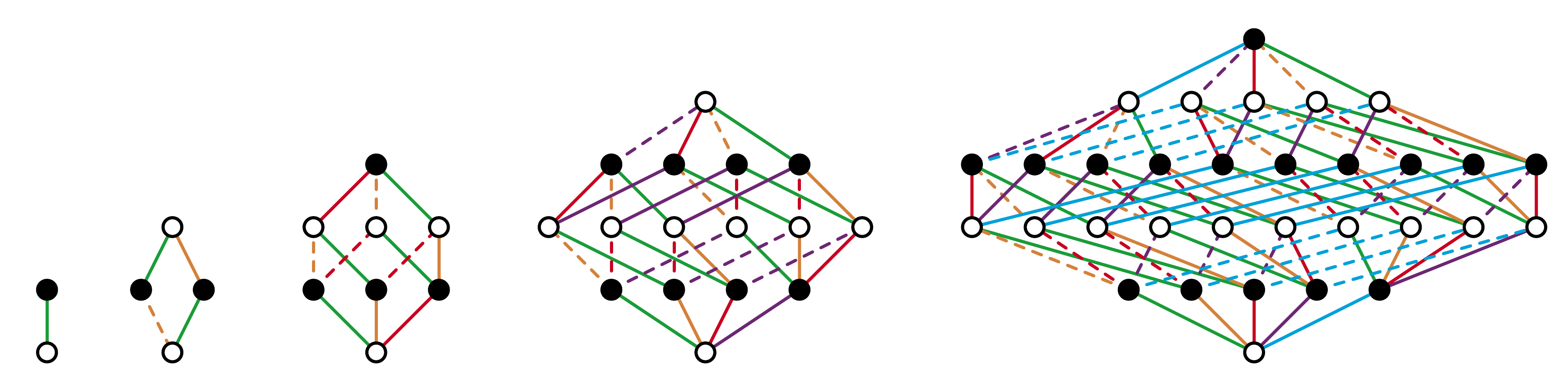}} \label{eXSeq}
\end{equation}
Besides these, Ref.\cite{r6-1} also identified the dash-chromotopology of the chiral and the twisted-chiral $N=4$ representations. These dash-chromotopologies differ only in the choice of edge-dashing and are equivalent to the two on the right-hand side of\eq{eN4split}. The chiral and twisted-chiral multiplets themselves are represented by Adinkras obtained from the two Adinkras on the right-hand side of\eq{eN4split} by raising one of the lowest scalar nodes to the top level in each.

However, Ref.\cite{r6-1} left open Conjecture~7.7: that a superfield of every dash-chromotopology can be somehow found, for Theorem~7.6 to construct from it superfield representations of all other supermultiplets of the same dash-chromotopology.

We prove in the next section that a superfield of every dash-chromotopology indeed exists, and provide an explicit construction for it.

\section{Trillions of Superfields}
 \label{10*12}
\begin{thrm}\label{t.7.7}
For every $N$ and every Adinkra dash-chromotopology, there exists a super-diffe\-ren\-ti\-ally constrained superfield describing an adinkraic supermultiplet with that chromotopology.
\end{thrm}
\begin{proof}
In superspace, the supersymmetry algebra\eq{eSuSy} is augmented by introducing the super-differential operators $D_I$, which satisfy:
\begin{equation}
  \{D_I,D_J\} = 2\,\delta_{IJ}\,H,\qquad
  [H,D_I]=0=\{Q_I,D_J\},\qquad I,J=1,\cdots,N.
 \label{eSuSyD}
\end{equation}
Acting on superfields, \ie, functions over superspace $(\t|\q^I)$, these operators admit a differential operator representation:
\begin{equation}
 D_I=\vd_I+i\d_{IJ}\,\q^J\,\ddt,\qquad
 Q_I=i\vd_I+\d_{IJ}\,\q^J\,\ddt,\qquad\text{where}\qquad
 \vd_I:=\pd{}{\q^I}.
 \label{eDQdiff}
\end{equation}
Consequently,
\begin{equation}
 D_I = -iQ_I+2i\d_{IJ}\,\q^J\,\ddt,\qquad\text{and}\qquad
 Q_I = iD_I+2\d_{IJ}\,\q^J\,\ddt.
 \label{eQ=D}
\end{equation}
Given a real, {\em\/a priori\/} unconstrained Salam-Strathdee superfield, $\IF$, its components are obtained by covariant projection\Ft{Brackets grouping indices denote weighted antisymmetrization: $A_{[I}B_{J]}:=\inv2(A_IB_J-A_JB_I)$, \etc\ The factors of $-i$ ensure reality of the components.}:
\begin{equation}
  \f:=\IF\big|,\quad
  \j_I:=-iD_I\IF\big|,\quad
   F_{[IJ]}:=iD_{[I}D_{J]}\IF\big|,\quad
   {\cal F}_{[I_1\cdots I_r]}:=(-i)^{\binom{r+1}2}D_{[I_1}\cdots D_{I_r]}\IF\big|,\cdots
  \label{eProjD}
\end{equation}
where the right-delimiting ``$|$'' denotes setting $\q^I\to0$. Since the $Q_I$'s and the $D_J$'s anticommute\eq{eSuSyD}, the projections\eq{eProjD}|and indeed any relationship written in terms of superfields and their $D$-derivatives|are covariant with respect to supersymmetry, generated by the $Q_I$'s.

For $\sC$ generated by the binary words $\bs{b}_a=(b_{a1},\cdots,b_{aN})$, and $\IF$ an {\em\/a priori\/} unconstrained Salam-Strathdee superfield, define
\begin{equation}
  \bs{b}_a\in\sC\quad\mapsto\quad D_1^{b_{a1}}\cdots D_N^{b_{a1}}
 \label{eDba}
\end{equation}
Owing to the anticommutivity of the distinct $D_I$'s, the monomials\eq{eDba} are in fact fully antisymmetric products.

To each code $\sC$ with a chosen set of $k$ generator codewords, there correspond $k$ super-differential monomials of the form\eq{eDba}. For doubly-even binary linear block codes, these super-differential monomials provide statistics-preserving maps between component fields, square to $+H^{\wt(\bs{b}_a)}$, and commute amongst each other.

The imposition of each one of the $k$ super-differential constraints
\begin{equation}
  \bs{b}_a\in\sC\quad\mapsto\quad
   \big[H^{\frac12\wt(\bs{b}_a)} + \vs_a D_1^{b_{a1}}\cdots D_N^{b_{a1}}\big]\,\IF=0,
    \qquad a=1,\cdots,k
 \label{eCba}
\end{equation}
halves the number of unrelated component fields in $\IF$. Owing to the mutual commutativity of the $D_1^{b_{a1}}\cdots D_N^{b_{a1}}$ monomials, these ``halvings'' may be applied jointly, resulting in a superfield where only $1/2^k$ of the initial components of the superfield $\IF$ remain unrelated. The relative signs $\vs_a$ in\eq{eCba} are the same ones from\eq{eM/C}.

However, the constraint system\eq{eCba} is not {\em\/strict\/}: the mappings provided by the operators $[H^{\frac12\wt(\bs{b}_a)} + \vs_aD_1^{b_{a1}}\cdots D_N^{b_{a1}}]$ are not a strict homomorphisms\cite{r6-3.2}, they leave behind certain ``orphan'' constants as remnants of almost completely eliminated component fields.

\paragraph{Example~1.} To illustrate this, consider the simplest, $N=4$ case with $\sC=d_4$. The super-differential constraint
\begin{equation}
  \big[H^2+D_1D_2D_3D_4\big]\IF=0,\qquad\text{for the choice }\vs=+1,
 \label{eD40}
\end{equation}
identifies, {\em\/via\/} Eqs.\eq{eProjD}:
\begin{gather}
  {\cal F}_{1234}=-\ddot\f,\qquad
  \dot\J_{IJK}=\ve_{IJK}{}^L\ddot\j_L,\qquad
  \ddot{F}_{[IJ]}=\inv{2!}\ve_{IJ}{}^{KL}\,\ddot{F}_{[KL]}.
 \label{eIds}
\end{gather}
These can be used to express almost all of ${\cal F}_{1234},\J_{IJK}$ and $F_{14},F_{24},F_{34}$ in terms of $\f,\j_I,F_{12},F_{13},F_{23}$, except for the constant term in $\J_{IJK}$'s and the constant and $\tau$-linear terms in $F_{14},F_{24},F_{34}$. The result is that the super-differential constraint system\eq{eD40} defines:
\begin{subequations}
 \label{eFD40}
\begin{gather}
  \IF|_{\text{(\ref{eD40})}}
   =\big(\f(\t)\big|\j_I(\t)\big|
           F_{12}(\t),F_{13}(\t),F_{23}(\t),f_{14}(\t),f_{14}(\t),f_{14}(\t)
                \big|\J_{[IJK]}(0)\big|0\big),\\
  \text{where}\quad
  f_{[I4]}(\t):= f_{[I4]}(0)+f'_{[I4]}(0)\t,\quad\text{for }I=1,2,3.
\end{gather}
\end{subequations}
The result\eq{eFD40} cannot be regarded an off-shell supermultiplet since the component fields $f_{[I4]}(\t)$ and $\J_{[IJK]}(0)$ satisfy $\t$-differential equations:
\begin{equation}
  \ddt^2\,f_{[I4]}~=~0~=~\ddt\,\J_{[IJK]}(0).
\end{equation}

To remedy this, note that the last group of identifications\eq{eIds} is suggestive: one really needs
\begin{equation}
  \ddot{F}_{[IJ]}=\inv{2!}\ve_{IJ}{}^{KL}\,\ddot{F}_{[KL]}
   \qquad\longrightarrow\qquad
  F_{[IJ]}=\inv{2!}\ve_{IJ}{}^{KL}\,F_{[KL]},
 \label{eSD}
\end{equation}
which is obtained, using the component projections\eq{eProjD}, as
\begin{equation}
  iD_{[I}D_{J]}\IF\big| = \inv2\ve_{IJ}{}^{KL}\,iD_{[K}D_{L]}\IF\big|.
\end{equation}
This then suggests replacing the super-differential condition\eq{eD40} with either of the two systems:
\begin{equation}
  \big[D_{[I}D_{J]} \mp \inv2\ve_{IJ}{}^{KL}D_{[K}D_{L]}\big]\IF^\pm=0,
   \qquad\textit{i.e.},\qquad
  \left\{\begin{aligned}{}
          [D_1D_2 \mp D_3D_4]\,\IF^\pm&=0,\\
          [D_1D_3 \pm D_2D_4]\,\IF^\pm&=0,\\
          [D_1D_4 \mp D_2D_3]\,\IF^\pm&=0.\\
         \end{aligned}\right.
 \label{eSysD4}
\end{equation}
Not surprisingly, this insures the full component field identification\eq{eSD}, and with both signs $F^\pm_{[IJ]}=\pm\inv{2!}\ve_{IJ}{}^{KL}\,F^\pm_{[KL]}$, rather than the weaker conditions\eq{eIds} insured by the single constraint\eq{eD40}.
 Next, applying $D_I$ on the system\eq{eSysD4} and evaluating at $\q^I\to0$ results in
$\J^\pm_{[IJK]} = \pm\ve_{IJK}{}^L\,\dot\j^\pm_L$, which again is precisely what is needed to fully eliminate $\J_{[IJK]}$ in terms of $\dot\j_I$, instead of the weaker identification\eq{eIds}.
 Finally, applying $D_{[I}D_{J]}$ on the system\eq{eSysD4} and evaluating at $\q^I\to0$ reproduces the final ${\cal F}^\pm_{1234}=\mp\ddot\f^\pm$.

This then leaves $(\f^\pm|\j^\pm_I|F^\pm_{12},F^\pm_{13},F^\pm_{23}|0|0)\subset (\f|\j_I|F_{IJ}|\J_{IJK}|\mathcal{F}_{1234})$, depicted as
\begin{equation}
   \IF|_{d_4,+}=\vC{\includegraphics[height=25mm]{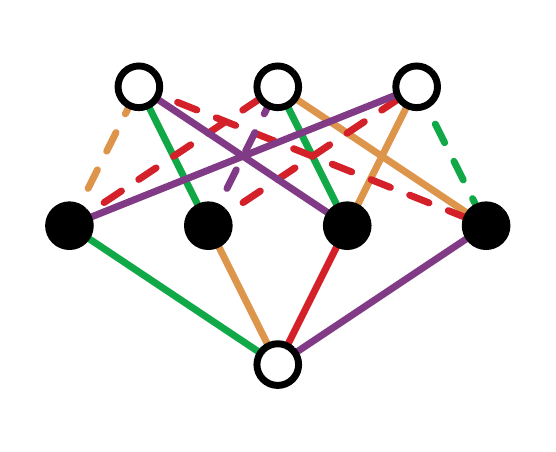}}
     \qquad\text{and}\qquad
   \IF|_{d_4,-}=\vC{\includegraphics[height=25mm]{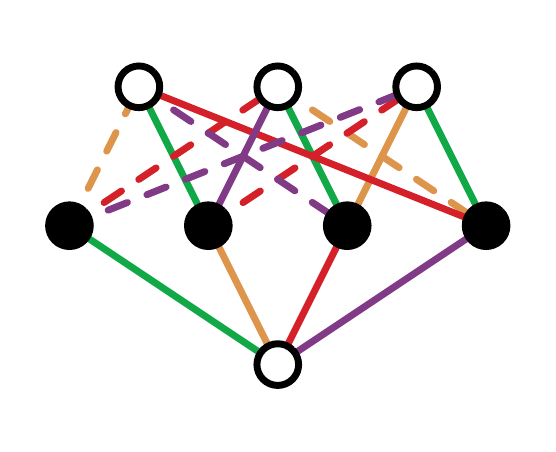}}
\end{equation}
spanning the two $d_4$-projected off-shell supermultiplets. The distinction between them is easily spotted: the product of signs along every four-colored quadrilateral equals the sign of the permutation of the colors in that quadrilateral in $\IF|_{d_4,+}$ and is opposite in $\IF|_{d_4,-}$. Also, both superfields are evidently sub-superfields of the {\em\/a priori\/} unconstrained $\IF$, depicted by the fourth Adinkra in the sequence\eq{eXSeq}.\hrulefill\vrule width.4pt height6pt depth0pt

It is thus the super-differential constraint\eq{eSysD4} rather than the na\"{\ii}ve\eq{eD40} that properly ``halves'' the $N=4$ real, {\em\/a priori\/} unconstrained superfield $\IF$. In turn, the equation\eq{eD40} may be regarded as the integrability condition for the system\eq{eSysD4}.%

The foregoing generalizes straightforwardly to all $N$ and all codes:

\begin{subequations}
 \label{eSFC}
\noindent{\small\sf\bfseries Construction~\ref{eSFC}}\hrulefill\vrule width.4pt height.4pt depth6pt
\begin{enumerate}\itemsep=-3pt\vspace{-3mm}
 \item Let $\IF$ be a real, {\em\/a priori\/} unconstrained Salam-Strathdee superfield.

 \item For every generator $\bs{b}_a$ of a code $\sC$, we define:
\begin{equation}
  {\cal I}(\bs{b}_a) := \{\,I=1,\cdots,N\mid b_{aI}=1\,\}.
\end{equation}
For example, ${\cal I}(110011)=\{1,2,5,6\}$ and ${\cal I}(101101)=\{1,3,4,6\}$.

 \item Associate to $\bs{b}_a$ the system  of $\inv2\binom{2w_a}{w_a}$ (anti)self-duality super-differential constraints:
\begin{equation}
 \Big\{\big[D_{[I_1}\cdots D_{I_{w_a}]}-
         \frc{\vs_a}{w_a!}\ve_{I_1{\cdots} I_{w_a}}{}^{J_1\cdots J_{w_a}}
           D_{[J_1}{\cdots} D_{J_{w_a}]}\big]\,\IF=0,\quad
            I_1,{\cdots},J_{w_a}\in{\cal I}(\bs{b}_a)\Big\},
 \label{eSysba}
\end{equation}
where $w_a:=\inv2\wt(\bs{b}_a)$ and $\vs_a=\pm1$, for all $a$.

 \item For every code generated by codewords $\{\bs{b}_1,\cdots,\bs{b}_k\}$, we impose a constraint system of the form\eq{eSysba}:
\begin{equation}
\begin{aligned}
  \IF|_{\sC,\vec\vs}:=
   \Big\{\IF:~&\big[D_{[I_1}\cdots D_{I_{w_a}]}
         -\frc{\vs_a}{w_a!}\ve_{I_1{\cdots} I_{w_a}}{}^{J_1\cdots J_{w_a}}
           D_{[J_1}{\cdots} D_{J_{w_a}]}\big]\,\IF=0,\\
              &~\text{for all }I_1,{\cdots},J_{w_a}\in{\cal I}(\bs{b}_a),~
                 \text{for each generator }\bs{b}_a\in\sC\Big\}.
\end{aligned}
 \label{eSysC}
\end{equation}
\hrulefill\vrule width.4pt height6pt depth0pt
\end{enumerate}
\end{subequations}

Each super-differential constraint system\eq{eSysba} has an {\em\/integrability\/} condition precisely of the form\eq{eCba}, where
\begin{equation}
  \frc12\big[ H^{\frac12\wt(\bs{b}_a)}+\vs_a D_1^{b_{a1}}\cdots D_N^{b_{aN}}\big]
 \label{eP}
\end{equation}
are a quasi-projection operators: for both $\vs_a=\pm1$, they square to a $H^{\frac12\wt(\bs{b}_a)}$-multiple of itself, and the two choices add up to $H^{\frac12\wt(\bs{b}_a)}\propto \ddt^{\frac12\wt(\bs{b}_a)}$. They are also in 1--1 correspondence with the code-generator projection operators of Refs.\cite{r6-3,r6-3.2}, which relates the two operators.

Finally, note that each super-differential constraint system\eq{eSysba} corresponding to each generator codeword of $\sC$ has precisely one relative sign, $\vs_a=\pm1$, stemming from \Eq{eM/C}. For $\sC$ being generated by $k$ codewords, the definition\eq{eSysC} churns out $2^k$ distinct superfields. Many of these may well be isomorphic, via a sign-redefinition on component fields.
 However, they do include all the inequivalent choices of edge-dashing in Adinkras, and so reproduce all the inequivalent dash-chromotopologies for Adinkras. Ref.\cite{r6-3.4} specifies a cohomology computation which tells if two given distinctly edge-dashed Adinkras are equivalent or not.
\end{proof}

\section{Examples and Conclusions}
 \label{s:C}
To illustrate the foregoing construction, we close with a few examples.

\paragraph{Example 2.~} Consider the next-simplest case of $\sC=d_6$, generated by $\bs{b}_1=(111100)$ and $\bs{b}_2=(001111)$. The super-differential constraint system\eq{eSysC} is now:
\begin{equation}
 \IF|_{d_6,(\vs_1,\vs_2)}:~\left\{\begin{aligned}{}
  [D_{[I}D_{J]}-\vs_1\inv2\ve_{IJ}{}^{KL}D_{[K}D_{L]}]\,\IF
    &=0,\quad I,J,K,L\in{\cal I}(111100)=\{1,2,3,4\},\\
  [D_{[I}D_{J]}-\vs_1\inv2\ve_{IJ}{}^{KL}D_{[K}D_{L]}]\,\IF
    &=0,\quad I,J,K,L\in{\cal I}(001111)=\{3,4,5,6\}.
                 \end{aligned}\right.
\end{equation}
Written out in full detail, this system becomes:
\begin{equation}
  \IF|_{d_6,(\vs_1,\vs_2)}:~
   \left\{~\begin{aligned}
  \big[D_1\,D_2-\vs_1D_3\,D_4\big]\IF&=0,\\
  \big[D_1\,D_3+\vs_1D_2\,D_4\big]\IF&=0,~
   \smash{\hbox{$\Bigg\}$}}~\text{for }\bs{b}_1=(111100),\\
  \big[D_1\,D_4-\vs_1D_2\,D_3\big]\IF&=0,\\[2mm]
  \big[D_3\,D_4-\vs_2D_5\,D_6\big]\IF&=0,\\
  \big[D_3\,D_5+\vs_2D_4\,D_6\big]\IF&=0,~
   \smash{\hbox{$\Bigg\}$}}~\text{for }\bs{b}_2=(001111),\\
  \big[D_3\,D_6-\vs_2D_4\,D_5\big]\IF&=0,\\[2mm]
          \end{aligned}\right.
 \label{eD6}
\end{equation}
Each of the two indicated groups of constraints independently halves the superfield $\IF$, so that jointly, they quarter it, from the initial $(1|6|15|20|15|6|1)$-dimensional representation to the minimal $(1|6|7|2)$-dimensional superfield, depicted by the Adinkra
\begin{equation}
  \IF|_{d_6,(--)}~=~\vC{\includegraphics[height=30mm]{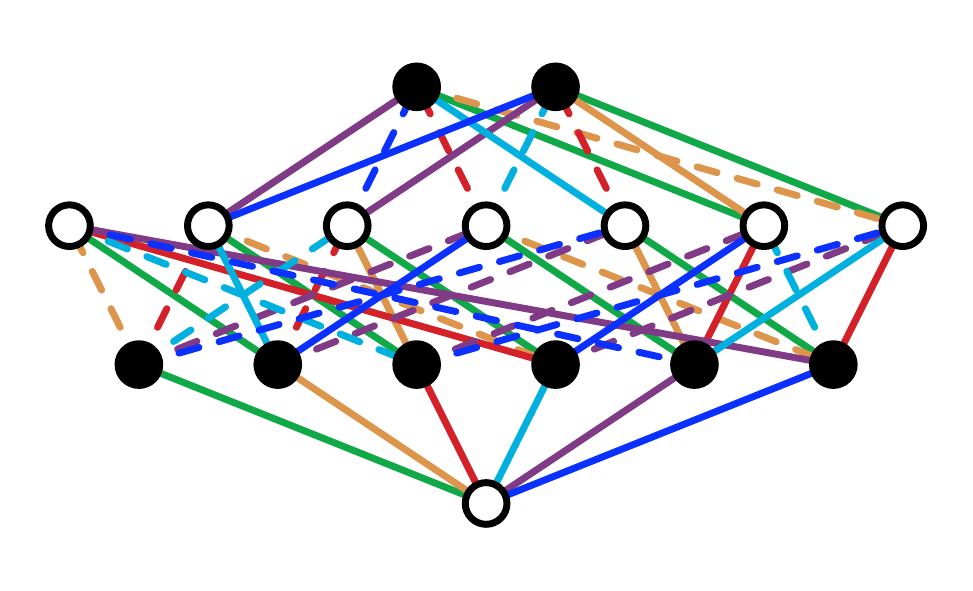}}
\end{equation}
The four different choices of signs, parametrized by $\vec\vs=(\pm1,\pm1)$, turn out to all yield choices of edge-dashing that are equivalent by field redefinition\cite{r6-3.4}, whence we show only one of them.\hrulefill\vrule width.4pt height6pt depth0pt

\paragraph{Example 3.~} Consider $\sC=h_8$, generated by $(11111111)$, and define the $N=8$ superfield:
\begin{equation}
  \IF|_{h_8,\vs}=
   \big\{\IF:~\big[D_{[I}D_JD_KD_{L]}
         -\vs\inv{4!}\ve_{IJKL}{}^{MNPQ}D_{[M}D_ND_PD_{Q]}\big]\IF=0\big\}.
 \label{eH8}
\end{equation}
The constraint system consists of a total of 35 equations; their {\em\/single\/} common integrability equation is $[H^4-\vs D_1\cdots D_8]\IF=0$. Jointly, they halve the original, $(1|8|28|56|70|56|28|8|1)$-dimensional representation to a $(1|8|28|56|35)$-dimensional superfield, representable by the Adinkra
\begin{equation}
  \vC{\begin{picture}(140,25)
       \put(-20,0){\includegraphics[height=27mm]{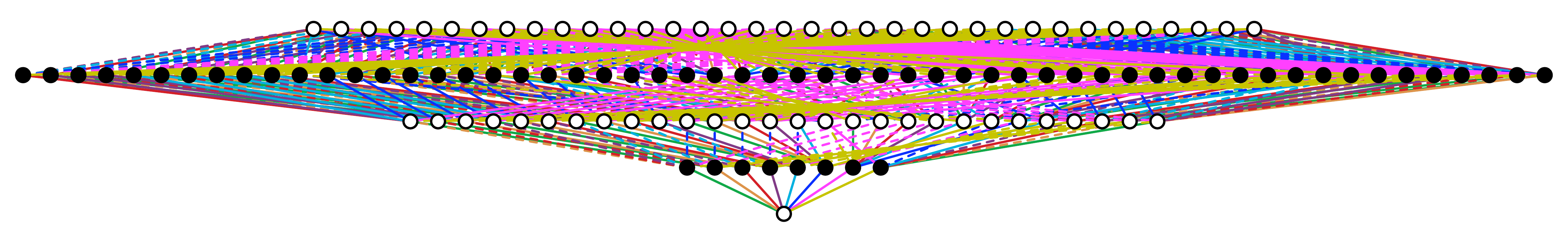}}
      \end{picture}}
\end{equation}
In this case, the two choices of the sign, $\vs=\pm1$, correspond to two inequivalent choices of edge-dashing\cite{r6-3.4}, but we omit the other Adinkra since their size and complexity obscures an easy spotting of the differences. Since $h_8$ is not maximal, this is not a minimal $N=8$ superfield.\hrulefill\vrule width.4pt height6pt depth0pt

\paragraph{Example 4.~} Finally, $\sC=e_8$ is generated by $\{(11110000),(00111100),(00001111),(01010101)\}$, and defines the $N=8$ superfield:
\begin{equation}
  \IF|_{e_8,\vec{\vs}}:~
   \left\{~\begin{array}{r@{\>}l@{\quad}r@{\>}r@{\>}l}
  \big[D_I\,D_J-\vs_1D_K\,D_L\big]\IF&=0, &I,J,K,L&\in{\cal I}(11110000)&=\{1,2,3,4\},\\
  \big[D_I\,D_J-\vs_2D_K\,D_L\big]\IF&=0, &I,J,K,L&\in{\cal I}(00111100)&=\{3,4,5,6\},\\
  \big[D_I\,D_J-\vs_3D_K\,D_L\big]\IF&=0, &I,J,K,L&\in{\cal I}(00001111)&=\{5,6,7,8\},\\
  \big[D_I\,D_J-\vs_4D_K\,D_L\big]\IF&=0, &I,J,K,L&\in{\cal I}(01010101)&=\{2,4,6,8\}.\\
          \end{array}\right.
 \label{eE8}
\end{equation}
This system consists of a total of 12 constraints; the integrability equation of each of the four indicated groups is of the form $[H^2-\vs_a D_ID_JD_KD_L]\IF=0$, with $I,J,K,L$ ranging over the corresponding four subsets ${\cal I}(\bs{b}_a)$, as specified in Eqs.\eq{eE8}. As a result, the $(1|8|28|56|70|56|28|8|1)$-dimensional {\em\/a priori\/} unconstrained superfield is chiseled down to a $(1|8|7)$-dimensional superfield, such as
\begin{equation}
  \IF|_{e_8,(--++)}~=~\vC{\includegraphics[height=30mm]{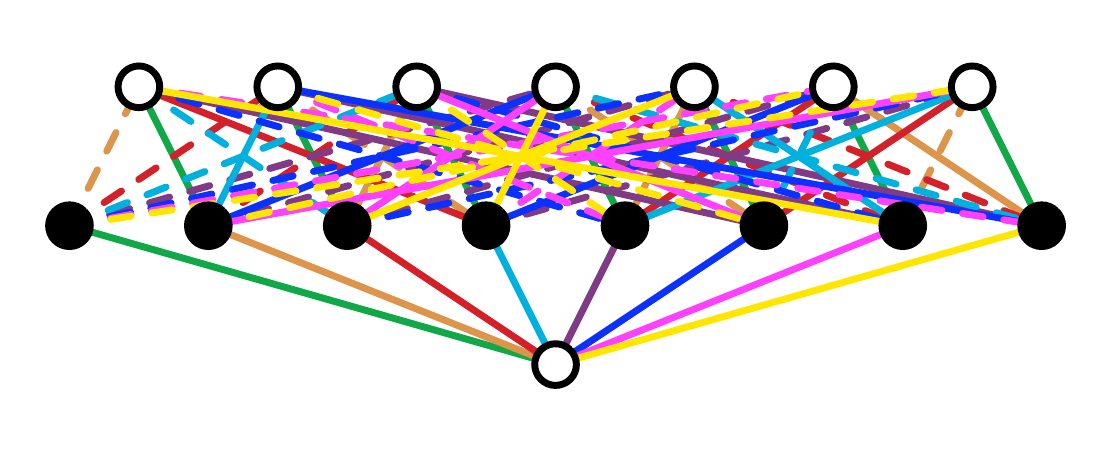}}
\end{equation}
which turns out to be closely related to the ``ultra-multiplet'' of Ref.\cite{rGR0}. The superfields\eq{eE8} are minimal. Noting that $(11110000)+(00001111)=(11111111)$, it follows that $h_8\subset e_8$, whereby $\IF|_{e_8,\vec\vs}\subset\IF|_{h_8}$. It is the combinatorial complexity of such embedding chains for $N>4$ that may be seen correlated with the surprising number of inequivalent supermultiplets\cite{r6-3,r6-3.2}.
\hrulefill\vrule width.4pt height6pt depth0pt

To summarize, we have presented a ``Construction~\ref{eSFC},'' which from
 \begin{enumerate}\itemsep=-3pt\vspace{-3mm}
 \item a real, {\em\/a priori\/} unconstrained Salam-Strathdee $N$-extended worldline superfield $\IF$,
 \item a doubly-even binary linear block code $\sC$ of length $N$ and with $k$ generators, and
 \item a $k$-tuplet of signs $\vec\vs$,
\end{enumerate}\vspace{-3mm}
custom-fashions a constrained sub-superfield $\IF|_{\sC,\vec\vs}\subset\IF$ with the $[0,1]^N/\sC$ chromotopology and the edge-dashing determined by $\vec\vs$. The collection of supermultiplets with all $\vec\vs$-choices include all inequivalent edge-dashings and we defer to Ref.\cite{r6-3.4} for the details of a cohomological computation that tells if two given $\vec\vs$-choices are equivalent or not, and how may inequivalent choices there exist.

Once we have the superfield $\IF|_{\sC,\vec\vs}$ resulting from Construction~\ref{eSFC}, the construction in Theorem~7.6 of Ref.\cite{r6-1} produces from $\IF|_{\sC,\vec\vs}$ every supermultiplet with the same dash-chromotopo\-logy. Counting all such superfields as different|after all, the supermultiplets they represent {\em\/are\/} conventionally considered different|the total count of so-obtained superfields (one for every Adinkra) is well beyond trillions\cite{r6-3,r6-3.2}. In another sense, for a given $N$ and a given chromotopology, Theorem~7.6 does effectively relate all superfields representing the differently ``hanged'' supermultiplets to one, such as the one obtained by Construction~\ref{eSFC}. In this sense, they are all related, whence the name ``family'' for their collection.

This situation is not quite as outlandish as it may seem: For example, it is well known that in 4-dimensional ${\cal N}=1$ supersymmetric spacetime, every chiral supermultiplet, $\F$, equals the super-derivative $D^2\IU$ of an {\em\/a priori\/} unconstrained, complex superfield $\IU$. Nevertheless, $\F$ and $\IU$ are regarded as different superfields for all practical purposes, and certainly provide inequivalent representations of supersymmetry.

In the same sense, the trillions or more of superfields defined by the use of Construction~\ref{eSFC} herein, Theorem~7.6 of Ref.\cite{r6-1}, the doubly-even binary linear block code classification\cite{r6-3,r6-3.2} and the cohomology computation of Ref.\cite{r6-3.4} are all just as different. Indeed, a comparison of the last two examples shows that $\IF|_{e_8,\vec\vs}\subset\IF|_{h_8}\subset\IF$ generalizes the relation $\F\subset\IU$ within 4-dimensional ${\cal N}=1$ supersymmetry. The combinatorial complexity of embedding chains for $N>4$ such as $\IF|_{e_8,\vec\vs}\subset\IF|_{h_8}\subset\IF$ may thus be seen as surprisingly large number of inequivalent supermultiplets\cite{r6-3,r6-3.2}. To this end, note also that a $\IF|_{h_8}$ generates, by way of Theorem~7.6, an entire family of supermultiplets and corresponding superfields, depicted by Adinkras that may be obtained from\eq{eH8} by hanging it from various subsets of nodes. The combinatorial complexity of this task|whence the enormous size of this resulting family|is evident, we trust.

The myriads of superfields obtainable by Construction~\ref{eSFC} are in many ways the higher-$N$, real analogues of $\F$, obtained with no symmetry assumed. Imposing symmetry relationships among the nodes evidently reduces the number of ways in which individual nodes can he raised or lowered. This then necessarily reduces the number of inequivalent Adinkras, superfields and supermultiplets: the bigger the additional symmetry requirements, the smaller the total number of inequivalent equivariant representations.

Of special interest are maximally projected, minimal supermultiplets, and all maximal codes usable to that end have been found\cite{r6-3}. It turns out that for $N<10$, such maximal codes and thus also the minimal supermultiplets are unique|but not so for $N\geq10$. For illustration, here are the two inequivalent minimal $N=10$ Adinkras:
\begin{gather}
  \sC=d_{10}: \quad\vC{\includegraphics[width=110mm]{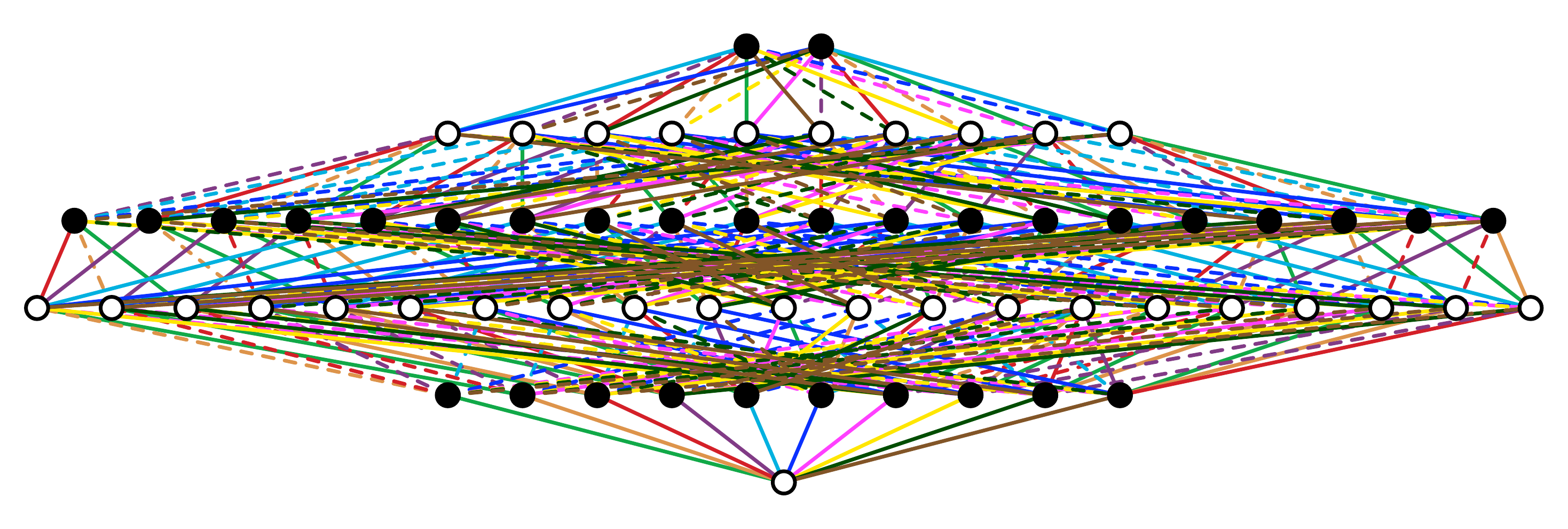}}    \label{D10}\\
  \sC=t_2+e_8:\quad\vC{\includegraphics[width=110mm]{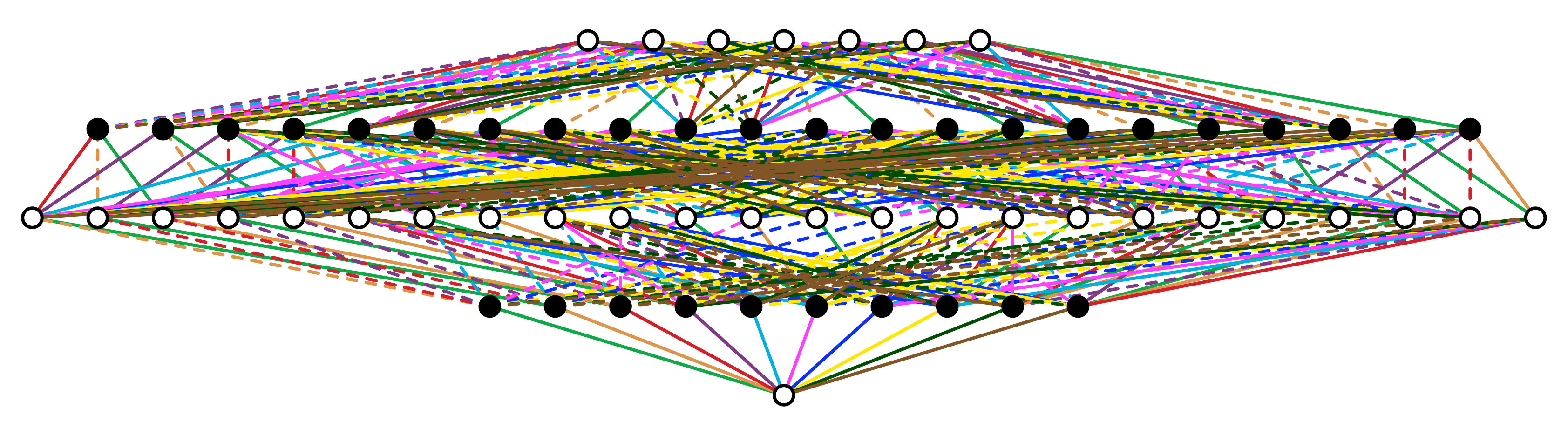}}  \label{I2xE8}
\end{gather}
and Construction~\ref{eSFC} produces a super-differentially constrained superfield for each. Already the count of component fields per engineering unit-level proves that they cannot be isomorphic. However, the superfields corresponding to the valise Adinkras of the respective chromotopologies|which Theorem~7.6 of Ref.\cite{r6-1} represents in terms of super-derivatives of the superfields\eqs{D10}{I2xE8}|turn out to be isomorphic\cite{r6-3.2}. In general, for $N\geq10$ there exist multiple minimal supermultiplets resulting from Construction~\ref{eSFC} and superfields|170 for $N=32$|but there will exist super-differential relations amongst them. We note in passing that the $N=16$ case also has two inequivalent minimal supermultiplets obtained by Construction~\ref{eSFC}, and which correspond to the codes $e_8\oplus e_8$ and $e_{16}$\cite{r6-3}, and which are in 1--1 correspondence with the 16-dimensional lattices $E_8{\times}E_8$ and $D_{16}$, respectively, and also the so-named Lie algebras.

Finally, this collection (trillions or so, for $N\leq32$) of superfields does not, by far, exhaust the listing of representations of $N$-extended worldline supersymmetry without central charges! Indefinitely more can be constructed by the usual methods of tensoring, (anti)symmetrizing and contracting|just as is the case with Lie algebras.

\begin{flushright}
 \sl
 \dots a brilliant diversity spread like stars,\\[-1mm]
 like a thousand points of light in a broad and peaceful sky.\\[-1mm]
 ---~William H.~Bush
\end{flushright}

\bigskip\paragraph{\bfseries Acknowledgments:}
This research was supported in part by the endowment of the John S.~Toll Professorship, the University of Maryland Center for String \& Particle Theory, National Science Foundation Grant PHY-0354401, and Department of Energy Grant DE-FG02-94ER-40854. The Adinkras were drawn with the aid of the {\em Adinkramat\/}~\copyright\,2008 by G.~Landweber.

\end{document}